\documentclass[aps,prl,showpacs,amsmath,amssymb,amsfonts,superscriptaddress,lengthcheck,twocolumn]{revtex4-1}

\usepackage{graphicx}
\usepackage{subfigure}
\usepackage{verbatim}
\usepackage{dcolumn}
\usepackage{bm}
\usepackage{epsf}
\usepackage{color}
\usepackage[colorlinks=true,citecolor=blue,linkcolor=blue,urlcolor=blue]{hyperref}
\usepackage{hhline}

\newcommand{\bra}[1]{\left\langle #1\right|}
\newcommand{\ket}[1]{\left|#1\right\rangle}

\newcommand{\tr}[1]{\mathrm{tr}\left\{#1\right\}}

\newcommand{\la}{\left\langle}
\newcommand{\ra}{\right\rangle}

\newcommand{\de}[1]{\delta\left(#1\right)}
\newcommand{\td}{\mathrm{d}}

\newcommand{\etal}{\textit{et al. }}
\newcommand{\e}[1]{\exp{\left(#1\right)}}
\newcommand{\lo}[1]{\ln{\left(#1\right)}}

\newcommand{\sh}[1]{\sinh{\left(#1\right)}}

\newcommand{\ct}[1]{\coth{\left(#1\right)}}
\newcommand{\bla}{bla\\bla\\bla\\bla\\bla}

\newcommand{\mc}[1]{\mathcal{#1}}

\newcommand{\mrm}[1]{\mathrm{#1}}

\DeclareMathOperator*{\sumint}{%
\mathchoice%
  {\ooalign{$\displaystyle\sum$\cr\hidewidth$\displaystyle\int$\hidewidth\cr}}
  {\ooalign{\raisebox{.14\height}{\scalebox{.7}{$\textstyle\sum$}}\cr\hidewidth$\textstyle\int$\hidewidth\cr}}
  {\ooalign{\raisebox{.2\height}{\scalebox{.6}{$\scriptstyle\sum$}}\cr$\scriptstyle\int$\cr}}
  {\ooalign{\raisebox{.2\height}{\scalebox{.6}{$\scriptstyle\sum$}}\cr$\scriptstyle\int$\cr}}
}

\begin{document}

\title{Quantum work and the thermodynamic cost of quantum measurements}

\author{Sebastian Deffner}
\affiliation{Theoretical Division, Los Alamos National Laboratory, Los Alamos, NM 87545, USA}
\affiliation{Center for Nonlinear Studies, Los Alamos National Laboratory, Los Alamos, NM 87545, USA}
\author{Juan Pablo Paz}
\affiliation{Departamento de F\'{i}sica, FCEyN, UBA, Ciudad Universitaria Pabell\'{o}n 1, 1428 Buenos Aires, Argentina}
\affiliation{IFIBA CONICET, FCEyN, UBA, Ciudad Universitaria Pabell\'{o}n 1, 1428 Buenos Aires, Argentina}
\author{Wojciech H. Zurek}
\affiliation{Theoretical Division, Los Alamos National Laboratory, Los Alamos, NM 87545, USA}

\date{\today}

\begin{abstract}
Quantum work is usually determined from two projective measurements of the energy at the beginning and at the end of a thermodynamic process.  However, this paradigm cannot be considered thermodynamically consistent as it does not account for the thermodynamic cost of these measurements. To remedy this conceptual inconsistency we introduce a novel paradigm that relies only on the expected change of the average energy given the initial energy eigenbasis. In particular, we completely omit quantum measurements in the definition of quantum work, and hence quantum work is identified as a thermodynamic quantity of only the system. As main results we derive a modified quantum Jarzynski equality and a sharpened maximum work theorem in terms of the information free energy. Comparison of our results with the standard approach allows to quantify the informational cost of projective measurements.
\end{abstract}


\maketitle

In classical mechanics and thermodynamics work is determined 
by a functional of a force along a trajectory in phase space
\cite{Callen1985a,
Boksenbojm2010,Deffner2013g}. For quantum 
systems the situation is much more subtle, 
since trajectories simply do not exist and a 
hermitian work operator  cannot be defined 
\cite{Talkner2007,Campisi2011}. Rather, 
quantum work is commonly determined as the difference
between two energies 
projectively measured
at two times: one at the beginning  and one at the end of 
a thermodynamic process \cite{Kurchan2000,Tasaki2000}. Despite the obvious limitations of this notion -- no truly open systems can be described since the change of internal energy comprises only work, but no heat -- the two-time energy measurement approach has proven to be practical and powerful. For instance, this approach has led to the experimental verification of quantum fluctuation theorems \cite{Batalhao2014,An2014b} and the development of heat engines at the nanoscale \cite{Abah2012a,Roßnagel2013,Dechant2014,Zhang2014a}.

Nevertheless, quantum work has remained somewhat elusive with many peculiar features and open questions \cite{Hanggi2015,Talkner2015}. For instance, it has only recently been pointed out that work can be measured at a single (final) time 
by means of a generalized measurement  \cite{Roncaglia2014a}, that
its probability distribution can be interferometrically estimated 
\cite{Vedral2012,Kafri2012,Venkatesh2013,Fusco2014a}, 
that it reduces to the classical notion of 
thermodynamic work in high-temperature \cite{Deffner2010a} 
and semi-classical  \cite{Jarzynski2015c} limits, and that the 
paradigm of two-time energy measurements  is also applicable 
to open systems as long as the dynamics is unital \footnote{A unital map is a completely positive, trace preserving map that also preserves the identity. Thermodynamically, a system evolving under unital dynamics  can be understood as an open system that is so weakly coupled that it only exchanges information, but no heat with the environment.}, see Refs.~\cite{Rastegin2013a,Albash2013,Rastegin2014,Deffner2015a,Deffner2015,Gardas2015,Manzano2015}.  

In the following we will motivate and introduce a new definition of 
quantum work based on a proper characterization of the role of quantum measurements -- a feature not present in the semi-classical limit \cite{Jarzynski2015c}. Imagine the typical situation for which the quantum Jarzynski equality is valid \cite{Deffner2013g}: at $t=0$ the system is prepared in a Boltzmann-Gibbs distribution, $\rho_0\propto \e{-\beta H_0}$, where $H_0$ is the initial Hamiltonian and $\beta$ denotes the inverse temperature. Then the density operator $\rho_0$ is diagonal in energy basis, and the projective measurement of the energy simply determines the thermal occupation probabilities $p(n_0)\propto\e{-\beta \epsilon(n_0)}$. After the first measurement, the system evolves under a time-dependent Hamiltonian $H(t)$ from $t=0$ to $t=\tau$, and the final state, $\rho_\tau$, is typically a complicated nonequilibrium state \cite{Deffner2008,Deffner2010,Deffner2013}. Generally, $\rho_\tau$ is not diagonal in the energy basis, and therefore a projective measurement of the energy is accompanied by a back-action on the system \cite{NielsenChuang}. In many realistic situations one does not have to worry too much about this back-action. Experimentally it poses a real challenge to fully isolate a quantum system from its environment and hence almost all, real quantum systems eventually decohere \cite{Brune1996,Zurek2003}. Interestingly, the energybasis can be  \textit{einselected} by the environment \cite{Paz1999}, and hence $\rho_\tau$ quickly becomes close to diagonal in energy eigenbasis. Hence, we believe that the methods put forward here will be also useful in the treatment of open systems, see e.g. also Refs.~\cite{Esposito2006}. For the time being, however, we will exclusively focus on isolated systems.

For fast processes -- faster than the time-scales over which decoherence happens -- or small systems $\rho_\tau$ is generally not diagonal, which has been seen explicitly, e.g., in the ion trap experiment by An \etal \cite{An2014b}. In such situations the back-action of the measurement on the system does play a role and results in an increase of von-Neumann entropy of the system, i.e., in a change of information. Information, however, is a thermodynamic resource \footnote{To the best of our knowledge the thermodynamic consequences of measurements have only been discussed for thermodynamic averages \cite{Jacobs2009,Jacobs2012,Kammerlander2016}, whereas the present focus is on fluctuation theorems.}, which can be related to a free energy usable to perfrom additional work \cite{Deffner2013c,Parrondo2015}. Hence, one would expect this additional (information) free energy to explicitly show up in the thermodynamic relations \cite{Jacobs2012,Yi2013}. However, in the standard treatment \cite{Campisi2011} and in experiments \cite{Batalhao2014,An2014b} this informational back-action on the system from the second projective measurement has not been considered. Hence, the paradigm of two-time measurements is thermodynamically incomplete.

In our analysis we will address this issue and resolve the conceptual inconstency arising from neglecting the informational contribution to the laws of thermodynamics. This will lead to a new notion of quantum work, which relies only on the internal energy as an average of the time-evolved energy eigenstates. We will see that this notion is in full agreement with the first law of thermodynamics -- the average work is given by the change of internal energy -- but that we can also quantify the informational contribution to the free energy from projective measurements. As a main result we will derive a modified quantum Jarzynski equality and an associated maximum work theorem, in which the thermodynamic free energy is replaced by the information free energy \cite{Deffner2012,Sivak2012,Still2012,Parrondo2015}. To the best of our knowledge no previous work has derived \emph{a modified Jarzynski equality and a sharper maximum work theorem} relating the cost of projective measurements with such \emph{a generalized free energy}.

\paragraph*{Two-time measurements -- notions and issues.}

We begin by briefly reviewing the paradigm of the two-time energy measurement approach, and establish notions and notations. Here and in the following we consider an isolated quantum system with time-dependent Schr\"odinger equation, $i\hbar\,|\dot{\psi}_t\rangle=H_t\,\ket{\psi_t}$. We are interested in describing thermodynamic processes that are induced by varying an external control parameter $\lambda_t$ during time $\tau$, with $H_t=H(\lambda_t)$. 

Commonly quantum work is determined by the following, experimentally motivated protocol: After preparation of the initial state $\rho_0$ a projective measurement of the energy is performed; then the system is allowed to evolve under the time-dependent Schr\"odinger equation, before a second projective energy measurement is performed at $t=\tau$. 

For the sake of simplicity and to avoid clutter in the formulas we further assume that the system is initially prepared in a Gibbs state, $\rho_0=\e{-\beta H_0}/Z_0$, where $\beta$ is the inverse temperature and $Z_0$ is the partition function, $Z_0=\tr{\e{-\beta H_0}}$. Strictly speaking the projective measurement at $t=0$ is superfluous for initially thermal states, since such $\rho_0$s are diagonal in the energy basis. The internal energy of the system, and therefore the full thermodynamic behavior, can simply be determined by an average over all energy eigenstates equipped with the thermal occupation probabilities \cite{Brito2014}, $p(n_0)=\e{-\beta\, \epsilon(n_0,\lambda_0)}/Z_0$. 

The first law of thermodynamics determines that the average work is given by the change of internal energy, $\la W\ra= \tr{\rho_\tau\,H_\tau}-\tr{\rho_0\,H_0}$, where $\rho_\tau=U_\tau \rho_0 U^\dagger_\tau$, and $U_{\tau}$ is the unitary time evolution operator, $U_{\tau}=\mc{T}_>\, \e{-i/\hbar\,\int_0^{\tau}\td t\,H_t}$, and $\mc{T}_> $ denotes time-ordering. Accordingly, for a single realization of the two-time measurement protocol the quantum work reads
\begin{equation}
\label{eq04}
W_{n_0\rightarrow n_\tau}=\epsilon(n_\tau,\lambda_{\tau})-\epsilon(n_0,\lambda_0)\,,
\end{equation}
where $\ket{n_0}$ is the initial eigenstate with eigenenergy $\epsilon(n_0,\lambda_0)$ and $\ket{n_\tau} $ with $\epsilon(n_\tau,\lambda_\tau)$ describes the final energy eigenstate.

The corresponding quantum work probability distribution \eqref{eq05} is then given by an average over 
an ensemble of realizations of this protocol, $\mc{P}(W)=\la \de{W-W_{n_0\rightarrow 
n_\tau}}\ra$, which can be written as \cite{Kafri2012,Leonard2014a}
\begin{equation}
\label{eq05}
\mathcal{P}(W)=\sumint_{n_0,n_\tau} \de{W-W_{n_0\rightarrow n_\tau}}\,
p\left(n_0; n_\tau\right).
\end{equation}
In the latter equation the symbol $\sumint$ denotes a sum over the discrete 
part of the eigenvalues spectrum and an integral over the continuous part. 
In Eq.~\eqref{eq05} $p\left(n_0; n_\tau\right)$ are the joint  
probabilities for detecting $n_0$ and $n_\tau$ in the two energy 
measurements \cite{Kafri2012,Leonard2014a,Deffner2015a}, 
\begin{equation}
\label{eq06a}
p\left(n_0; n_\tau\right)= \tr{\Pi_{n_\tau} U_\tau \Pi_{n_0}\rho_0 \Pi_{n_0} U_\tau^\dagger}\,,
\end{equation}
where, $\Pi_n$ denotes the projector into the space spanned by the $n$th eigenstate, which becomes for non-degenerate spectra $\Pi_n=\ket{n}\bra{n}$. It is then a simple exercise to show that from the definition 
of $\mc{P}(W)$ \eqref{eq05} we have the quantum Jarzynski 
equality \cite{Talkner2007,Campisi2011},
\begin{equation}
\begin{split}
\label{eq06}
\la e^{-\beta W}\ra=\int\td W\, \mc{P}(W)\,e^{-\beta W}=e^{-\beta \Delta F}\,,
\end{split}
\end{equation}
where $\Delta F=F_\tau-F_0$ and $F_t=-(1/\beta)\,\lo{Z_t}$. 

\paragraph*{Neglected informational cost.}

Generally the final state $\rho_\tau$ is a complicated nonequilibrium state. This means, in particular, that $\rho_\tau$ 
does not commute with the final Hamiltonian $H_\tau$, and one has to consider the back-action on the system due to the projective measurement of the energy \cite{NielsenChuang}. For a single measurement, $\Pi_{n_\tau}$, the post-measurement state is given by $\Pi_{n_\tau} \rho_\tau\Pi_{n_\tau}/p_n$, where $p_n=\tr{\Pi_{n_\tau}\,\rho_\tau}$. Thus, the system can be found on average in
\begin{equation}
\label{eq07}
\rho^M_\tau=\sum_{n_\tau}\Pi_{n_\tau}\,\rho_\tau\,\Pi_{n_\tau}\,.
\end{equation}
 Accordingly, the final measurement of the energy is accompanied by a change of information, i.e., by a change 
of the von Neumann entropy of the system
\begin{equation}
\label{eq08}
\Delta \mc{H}^M=-\tr{\rho^M_\tau\lo{\rho^M_\tau}}+\tr{\rho_\tau\lo{\rho_\tau}}\geq 0\,.
\end{equation}
Information, however, is physical \cite{Landauer1991} and its acquisition ``costs'' work. 
This additional work has to be paid by the external observer -- the measurement 
device. In a fully consistent thermodynamic framework this cost has to 
be taken into consideration \cite{Deffner2013c}, in particular when calculating 
the efficiency of thermodynamic devices \cite{Abah2012a,Roßnagel2013}.

\paragraph*{Quantum work without measurements.}

To remedy this conceptual inconsistency arising from neglecting the informational contribution of the projective measurements, we propose an alternative paradigm. For isolated systems quantum work is clearly given by the change of internal energy. As a statement of the first law of thermodynamics this holds true no matter whether the system is measured or not. Quantum measurements, however, can be understood as an interaction with a ``measuring environment''. Moreover, almost any environment induces decoherence \cite{Zurek2003}. Thus, defining work with the help of the environment and ignoring the effect of decoherence is as thermodynamically inconsistent as defining work via an external measurement and neglecting the informational cost of these projective measurements. 

For thermal states measuring the energy is superfluous as state and energy commute. Hence, a notion of quantum work can be formulated that is fully based on the time-evolution of energy eigenstates \footnote{The thermal state is then emulated \cite{Brito2014} by equipping each energy eigenstate  with the thermal occupation probabilities, $p(n_0)=\e{-\beta\, \epsilon(n_0,\lambda_0)}/Z_0$.}. Quantum work for a single realization is then determined by considering 
how much the expectation value for a single energy eigenstate changes 
under the unitary evolution \footnote{It is interesing to note that a similiar notion of quantum work has been considered in a different context in Refs.~\cite{Allahverdyan2005,Campisi2013}.}. Hence, we define
\begin{equation}
\label{eq09}
\widetilde{W}_{n_0}\equiv \bra{n_0} U^\dagger_\tau\,H_\tau\,U_\tau\ket{n_0}-\epsilon(n_0,\lambda_0)\,.
\end{equation}
We can easily verify that the so defined quantum work \eqref{eq09}, indeed, fulfills the first law. To this end, we compute the average work $\la W\ra_{\widetilde{\mc{P}}}$ for the modified quantum work distribution $\widetilde{\mc{P}}(W)$, and we obtain,
\begin{equation}
\label{eq11}
\begin{split}
\la W\ra_{\widetilde{\mc{P}}}&=\sumint_{n_0} \bra{n_0} U^\dagger_\tau\,H_\tau\,U_\tau\ket{n_0}\,p(n_0)-\tr{\rho_0\, H_0}\\
&=\tr{\rho_\tau\,H_\tau}-\tr{\rho_0\,H_0}=\la W\ra \,.
\end{split}
\end{equation}
It is important to note that the average quantum work determined from two-time energy measurements is identical to the (expected) value given only knowledge from a single measurement at $t=0$.  Most importantly, however, in our paradigm the external observer does not have to pay a thermodynamic cost associated with the change of information due to measurements. Hence, the present paradigm can be considered thermodynamically consistent and complete. 

\paragraph*{Modified quantum Jarzynski equality.}

What we have seen so far is that the first law of thermodynamics is immune to whether the energy of the system is measured or not, since projective measurements of the energy do not affect the internal energy. However, the informational content of the system of interest, i.e., the entropy, crucially depends on whether the system is measured. Therefore, we expect that the statements of the second law have to be modified to reflect the informational contribution \cite{Deffner2013c}. In our paradigm the modified quantum work distribution becomes
\begin{equation}
\label{eq10}
\widetilde{\mc{P}}(W)=\sumint_{n_0} \delta(W-\widetilde{W}_{n_0})\, p(n_0)\,,
\end{equation}
where as before $p(n_0)=\e{-\beta\, \epsilon(n_0,\lambda_0)}/Z_0$. Now, we can compute the average exponentiated work,
\begin{equation}
\label{eq12}
\la e^{-\beta W}\ra_{\widetilde{\mc{P}}}=\frac{1}{Z_0}\,\sumint_{n_0}\,e^{-\beta \,\bra{n_0} U^\dagger_\tau\,H_\tau\,U_\tau\ket{n_0}}\,.
\end{equation}
The right side of Eq.~\eqref{eq12} can be interpreted as the ratio 
of two partition functions, where $Z_0$ describes the initial thermal 
state. The second partition function 
$\widetilde{Z}_\tau\equiv \sumint_{n_0}\,\e{-\beta \,\bra{n_0} U^\dagger_\tau\,H_\tau\,
U_\tau\ket{n_0}}$ corresponds to the best possible guess for a thermal 
state of the final system given only the time-evolved energy eigenbasis. 
This state can be written as
\begin{equation}
\label{eq13}
\widetilde{\rho}_\tau\equiv \frac{1}{\widetilde{Z}_\tau}\,\sumint_{n_0}e^{-\beta \,\bra{n_0} U^\dagger_\tau\,H_\tau\,U_\tau\ket{n_0}}\,U_\tau\ket{n_0}\bra{n_0}U^\dagger_\tau,
\end{equation}
which differs from the true thermal state, $\rho_\tau^\mrm{eq}=\e{-\beta H_\tau}/Z_\tau$.

In information theory the ``quality'' of such a best possible guess is quantified 
by the relative entropy \cite{Kullback1954,Umegaki1954}, which measures the 
distinguishability of two (quantum) states. Hence, let us consider
\begin{equation}
\label{eq14}
S(\widetilde{\rho}_\tau||\rho_\tau^\mrm{eq})=\tr{\widetilde{\rho}_\tau\lo{\widetilde{\rho}_\tau}}-\tr{\widetilde{\rho}_\tau\lo{\rho_\tau^\mrm{eq}}}\,,
\end{equation}
for which we compute both terms separately. For the first term, the negentropy of $\widetilde{\rho}_\tau$ we obtain,
\begin{equation}
\label{eq15}
\begin{split}
&\tr{\widetilde{\rho}_\tau\lo{\widetilde{\rho}_\tau}}=-\ln{(\widetilde{Z})}\\
&-\beta\, \tr{\widetilde{\rho}_\tau \sumint_{m_0} \bra{m_0} U^\dagger_\tau\,H_\tau\,U_\tau\ket{m_0}\,U_\tau\ket{m_0}\bra{m_0}U^\dagger_\tau }\\
&=-\ln{(\widetilde{Z})}-\beta \widetilde{E}\,,
\end{split}
\end{equation}
where we introduced the expected value of the energy, $\widetilde{E}$, under the time-evolved eigenstates,
\begin{equation}
\label{eq16}
\widetilde{E}=\frac{1}{\widetilde{Z}} \sumint_{n_0}e^{-\beta \,\bra{n_0} U^\dagger_\tau\,H_\tau\,U_\tau\ket{n_0}}\bra{n_0} U^\dagger_\tau\,H_\tau\,U_\tau\ket{n_0}\,.
\end{equation}
The second term of Eq.~\eqref{eq14}, the cross entropy of $\widetilde{\rho}_\tau $ and $\rho_\tau^\mrm{eq} $, simplifies to
\begin{equation}
\label{eq17}
\begin{split}
&\tr{\widetilde{\rho}_\tau\lo{\rho_\tau^\mrm{eq}}}=-\lo{Z_\tau}\\
&-\beta \,\tr{\sumint_{n_0}\frac{1}{\widetilde Z} e^{-\beta \,\bra{n_0} U^\dagger_\tau\,H_\tau\,U_\tau\ket{n_0}} U\ket{n_0}\bra{n_0}U^\dagger\,H_\tau}\\
&=-\lo{Z_\tau}-\beta \widetilde{E}\,.
\end{split}
\end{equation}
Hence, the modified quantum Jarzynski equality \eqref{eq12} becomes
\begin{equation}
\label{eq18}
\la e^{-\beta W}\ra_{\widetilde{\mc{P}}}=e^{-\beta \Delta F}\,e^{-S(\widetilde{\rho}_\tau||\rho_\tau^\mrm{eq})}\,,
\end{equation}
where as before $\Delta F=-1/\beta\, \lo{Z_\tau/Z_0}$. Jensen's inequality further implies,
\begin{equation}
\label{eq19}
\beta\,\la W\ra \geq \beta\,\Delta F + S(\widetilde{\rho}_\tau||\rho_\tau^\mrm{eq})
\end{equation}
where we used $\la W\ra_{\widetilde{\mc{P}}}=\la W\ra$ \eqref{eq11}. 

Equations~\eqref{eq18} and \eqref{eq19} are our main results. By defining quantum work as an average over time-evolved eigenstates we obtain a modified quantum Jarzynski equality \eqref{eq18} and a generalized maximum work theorem \eqref{eq19}, in which the thermodynamic cost of projective measurements becomes apparent. These results become even more transparent by noting that similar versions of the maximum work theorem have been derived in the thermodynamics of information \cite{Vaikuntanathan2009,Deffner2012,Deffner2013c,Parrondo2015}. In this context it has proven useful to introduce the notion of an information free energy, 
\begin{equation}
\label{eq20}
\widetilde{F}_\tau=F_\tau+S(\widetilde{\rho}_\tau||\rho_\tau^\mrm{eq})/\beta\,.
\end{equation}
This free energy is a true thermodynamic quantity \cite{Deffner2012,Parrondo2015} that accounts for the additional capacity of a thermodynamic system to perform work due to information \cite{Deffner2013c}. Note that in the present context $\widetilde{F}_\tau$ is computed for the fictitious thermal state $\widetilde{\rho}_\tau$ \eqref{eq13}, whereas one usually considers the information free energy for the nonequilibirum state $\rho_\tau$ \cite{Vaikuntanathan2009,Deffner2012,Deffner2013c,Parrondo2015}. 

We can rewrite Eq.~\eqref{eq19} as
\begin{equation}
\label{eq21}
\beta\,\la W\ra \geq \beta\,\Delta \widetilde{F}\,.
\end{equation}
 Equation~\eqref{eq21} constitutes a sharper bound than the usual maximum work theorem, and it accounts for the extra free energy available to the system. Free energy, however, describes the usable, extractable work. In real-life applications one is more interested in the maximal free energy the system has available, then in the work that could be extracted by intermediate, disruptive measurements of the energy. Therefore, our treatment could be considered thermodynamically more relevant than the two-time measurement approach.

\paragraph*{Illustrative example: parametric harmonic oscillator.}

For the remainder of this discussion we will turn to an analytically solvable example, namely the parametric harmonic oscillator,
\begin{equation}
\label{eq21a}
H=p^2/2m+m \omega_t^2\,x^2/2\,.
\end{equation}
This system has been studied extensively in the literature, and it can be shown that the average work performed by changing the angular frequency from $\omega_0$ to $\omega_\tau$ is given by \cite{Deffner2008,Deffner2010a}
\begin{equation}
\label{eq22}
\la W \ra =\hbar/2\,\left(Q^* \omega_\tau-\omega_0\right)\,\ct{\beta/2\,\hbar\omega_0}\,.
\end{equation}
The quantity $Q^*$ is a measure of adiabaticty \cite{Husimi1953a,Deffner2010a}, which fully encodes the dynamics. In particular, we have $Q^*=1$ for adiabatic, infinitely slow processes, and $Q^*>1$ for finite time driving. The change in equilibrum free energy becomes \cite{Deffner2008}
\begin{equation}
\label{eq23}
\Delta F=\frac{1}{\beta}\,\lo{\frac{\sh{\beta/2\,\hbar\omega_\tau}}{\sh{\beta/2\,\hbar\omega_0}}}\,.
\end{equation}
Therefore, we merely have to compute the partition function $\widetilde{Z}$, which can be written as
\begin{equation}
\label{eq24}
\widetilde{Z}_\tau=\sum_{n_0}\e{-\sum_{n_\tau}\beta\hbar\omega_\tau\left(n_\tau+1/2\right)p_{n_0,n_\tau}}\,,
\end{equation}
where $p_{n_0,n_\tau}=\tr{\Pi_{n_\tau}\, U_{\tau}\, \Pi_{n_0}\, U_{\tau}^\dagger}$. Hence, $\widetilde{Z}$ is fully determined by the average final occupation number \cite{Deffner2008}
\begin{equation}
\label{eq25}
\la n_\tau\ra_{n_0}=\sum_{n_\tau} n_\tau p_{n_0,n_\tau}=\left(n_0+1/2\right) Q^*-1/2\,,
\end{equation}
from which we obtain
\begin{equation}
\label{eq26}
\begin{split}
\widetilde{Z}_\tau&=\sum_{n_0}\e{-\beta \hbar\omega_\tau Q^* (n_0+1/2)}\\
&= \left(2 \sh{\beta/2 \,Q^*\,\hbar\omega_\tau}\right)^{-1}\,.
\end{split}
\end{equation}
Accordingly, the informational correction to the maximum work theorem arising from omitting the second projective measurement becomes
\begin{equation}
\label{eq27}
S(\widetilde{\rho}_\tau||\rho_\tau^\mrm{eq})=\lo{\frac{\sh{\beta/2 \,Q^*\,\hbar\omega_\tau}}{\sh{\beta/2 \,\hbar\omega_\tau}}}\,,
\end{equation}
which is clearly non-negative and a simple function of the measure of adiabaticity $Q^*$. Note that for adiabatic processes, $Q^*=1$, the information free energy \eqref{eq19} becomes identical to the equilibrium free energy, since we have $S(\widetilde{\rho}_\tau||\rho_\tau^\mrm{eq})=0$.

\begin{figure}
\includegraphics[width=.48\textwidth]{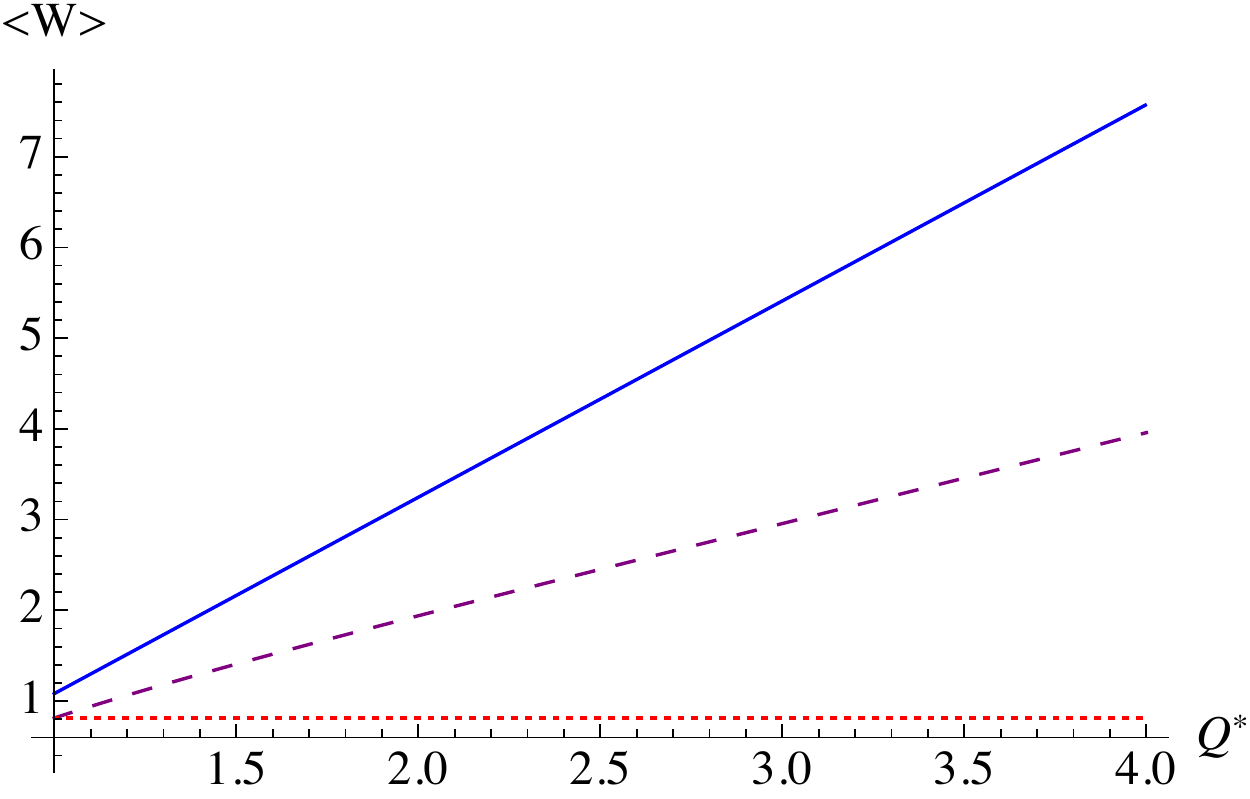}
\caption{\label{fig1} (color online) Average work $\la W\ra$ \eqref{eq22} (blue, solid line), together with the change of equilibrium free energy $\Delta F$ \eqref{eq23} (red, dotted line) and the informational maximum work theorem, $\beta \Delta F+S(\widetilde{\rho}_\tau||\rho_\tau^\mrm{eq})$ \eqref{eq19} (purple, dashed line) for the parametric harmonic oscillator \eqref{eq21a} with $\hbar=1$, $\beta=1$, $\omega_0=1$, and $\omega_\tau=2$.}
\end{figure}

\begin{figure}
\includegraphics[width=.48\textwidth]{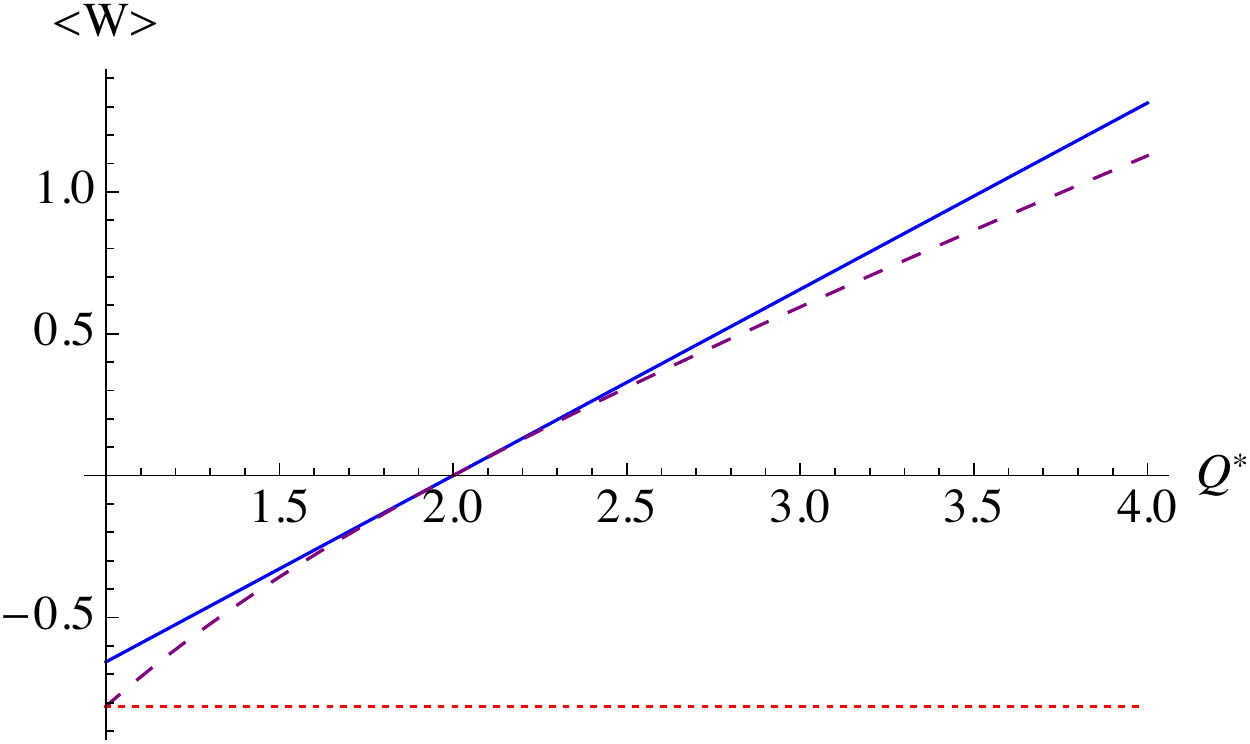}
\caption{\label{fig2} (color online) Average work $\la W\ra$ \eqref{eq22} (blue, solid line), together with the change of equilibrium free energy $\Delta F$ \eqref{eq23} (red, dotted line) and the informational maximum work theorem, $\beta \Delta F+S(\widetilde{\rho}_\tau||\rho_\tau^\mrm{eq})$ \eqref{eq19} (purple, dashed line) for the parametric harmonic oscillator \eqref{eq21a} with $\hbar=1$, $\beta=1$, $\omega_0=2$, and $\omega_\tau=1$.}
\end{figure}

In Figs.~\ref{fig1} and \ref{fig2} we plot the average work $\la W\ra$ \eqref{eq22} together with the change of equilibrium free energy $\Delta F$ \eqref{eq23} and the modified maximum work theorem \eqref{eq19}. We observe that the bound arising from the information free energy \eqref{eq20} is process dependent and sharper -- for some parametrizations even tight. 

\paragraph*{Concluding remarks.}

A conceptually consistent and complete framework of quantum thermodynamics crucially depends on accounting for quantum features and peculiarities \cite{Gemmer2009a}.  Indeed, one can hope that a completely quantum approach based on symmetries of entanglement \cite{DeffnerZurek2016} can help resolve outstanding problems of the classical and quantum points of view. In the present analysis we have shown that despite its success the two-time energy measurement approach to quantum work neglects the informational back-action of the projective measurements. This informational contribution to the laws of quantum thermodynamics has been highlighted by introducing a new paradigm in which quantum work is fully determined by the change of internal energy as an average over the initial energy eigenstates. This approach has allowed us to derive a modified quantum Jarzynski equality and a modified maximum work theorem, in which the equilibrium free energy is replaced by the information free energy. In conclusion, we achieved several important insights: (i) we have proposed a thermodynamically consistent notion of quantum work, which does not rely on external observers and projective measurements and (ii) we have included the thermodynamic cost of information gain in the paradigm of quantum work, and hence taken an instrumental step towards a conclusive theory of quantum thermodynamics of quantum information.

\acknowledgements{It is a pleasure to thank Jordan M. Horowitz, Christopher Jarzynski, Jim Crutchfield, and Gavin Crooks for interesting and insightful discussions. SD acknowledges financial support by the U.S. Department of Energy through a LANL Director's Funded Fellowship, and WHZ acknowledges partial support by the Foundational Questions Institute grant \# 2015-144057 on ``Physics of What Happens''.}

\bibliography{one_time_energy_revised}

\end{document}